\documentstyle[12pt]{article} 

\begin{document} 

\thispagestyle{empty} 

\vspace*{5mm} 

\begin{center} 
{\LARGE The London field in bulk} 

\smallskip 

{\LARGE layered superconductors} 

\vspace{12mm} 

{\large Giovanni Modanese} 
\footnote{e-mail: modanese@science.unitn.it} 

\medskip 

{\em European Centre for Theoretical Studies in Nuclear Physics 
and Related Areas \par 
Villa Tambosi, Strada della Tabarelle 286 \par 
I-38050 Villazzano (TN) - Italy} 

\medskip 

{and} 

\medskip 

{\em I.N.F.N.\ -- Gruppo Collegato di Trento \par 
Dipartimento di Fisica dell'Universit\`a \par 
I-38050 Povo (TN) - Italy} 

\end{center} 

\vspace*{5mm} 

\begin{abstract} 

In a composite superconductor in uniform rotation, the London field 
strength at equilibrium is given by the usual expression $B_L = 
2m\omega/e$; here $m$ corresponds to the bare electron mass, although the 
effective mass $m^*$ can be different in different layers. In the presence 
of strong angular accelerations, however, transient phases and differences 
in $m^*$ can become relevant. In particular, if the layers are orthogonal 
to the rotation axis and one of them is much thinner than the others, the 
superconductor may enter metastable states characterized by intense 
``slide" supercurrents in this layer. We investigate this phenomenon 
theoretically and with reference to current experiments.

\medskip

\noindent 74.25.Ha Magnetic properties 

\noindent 74.60.Jg Critical currents 

\noindent 74.62.Dh Effects of crystal defects, doping and substitution 

\noindent 74.80.-g Spatially inhomogeneous structures 

\noindent 74.80.Dm Superconducting layer structures: superlattices, 
heterojunctions, and multilayers 


\bigskip 

\end{abstract} 

\subsection*{1. Introduction} 

It is known that inside a rotating superconductor (SC) the magnetic field 
does not vanish, but takes a certain value $B_L$ called ``London field 
strength"; this is given, in terms of the electron charge $e$, the 
electron mass $m$ and the angular velocity $\omega$ of the SC, by a very 
simple expression \cite{London,Waldram}, namely 
\begin{equation} 
B_L = 2m\omega/e 
\label{e1} 
\end{equation} 
This relation holds for any value of the external magnetic 
field, including the case $B_{ext}=0$. The strength of $B_L$ is determined 
by the London equation (or better by a generalization of it---see 
\cite{Liu}), written in the reference frame co-rotating with the SC. 

The London field can be interpretated as the field needed to give the 
superelectrons the same rotation velocity of the crystal lattice. While 
the normal electrons keep pace with the positive ions due to ohmic 
friction, the Cooper pairs are mechanically decoupled from the lattice and 
need the field $B_L$ to move on circular orbits with frequency $\omega$. 
This interpretation also justifies a simple intuitive argument 
\cite{Anderson} to explain {\it a posteriori} the presence of the London 
field inside a rotating SC, with strength given by (\ref{e1}): it is 
essential that over the sample as a whole $v_{pairs}=v_{lattice}$, because 
otherwise very large bulk currents would flow. 

Starting in the Sixties, the London field has been measured in several 
experiments. In all measurements it was found that the mass $m$ defining 
the strength $B_L$ corresponds to the bare electron mass, i.e., not the 
effective mass $m^*$, renormalized by the interactions with the lattice, 
but the mass of the free electron (apart from small relativistic 
corrections \cite{Gutfreund} and some further minor corrections 
\cite{Peskin}). This made possible to use precise measurements of $B_L$ to 
deduce precise values of the parameters $e$ and $m$ (and also $\hbar$, 
since in a rotating ring of area $S$ the flux is quantized and the quanta 
correspond to steps $\Delta \nu = \hbar/4mS$ in the rotation frequency) 
\cite{Exps}. 

The easiest way to check that $m$ is the bare electron mass and not the 
effective mass is by starting from the full Hamiltonian of the solid 
including the electrons, the ions, and all their interactions. 
Transforming to the rotating frame, one finds that each electron ``feels" 
the additional vector potential 
\begin{equation} 
{\bf A}_{\bf \omega}({\bf r}) = m/e ({\bf \omega} \times {\bf r}) 
\label{e2} 
\end{equation} 
(with field $(m/e){\bf \omega}$), regardless of its complicated 
interaction with the ions and the other electrons.

\subsection*{2. Acceleration-deceleration phases. Variations of $m^*$ in a 
layered disk} 

Within this well-established framework, two important variations can 
occur. Let us consider cases when 

\medskip 

(i) The rotation velocity of the SC is suddenly changed 

\medskip 

or 

\medskip 

(ii) The SC has a composite structure, being made of two or more parts 
with different chemical and crystal properties 
\footnote{It is known for instance that the crystal structure of YBCO 
depends on the oxigen doping of the material. It happens quite frequently 
that the doping process of large ceramic samples results in portions of 
the samples having different oxigen content. The effective electron mass 
$m^*$ (typically 4-5 times larger than the bare mass) depends on this 
content. Compare also Section 4.} 
and different values of the effective electron mass $m^*$, 

\medskip 

or both (i) and (ii). 

\medskip 

As we shall show in the following, the combined effect of 
accelerations-decelerations and inhomogeneities 
in the material can spoil the dynamical 
equilibrium between superelectrons and lattice ions usually associated 
with the London field. 

When this happens, the relative velocity of the superelectrons with 
respect to the lattice ions can indeed lead to strong ``slide" 
supercurrents---although we shall see that these are not bulk currents, 
but are always confined to thin layers. 

It is quite straightforward to evaluate the order of magnitude of these 
currents. Let us consider a ceramic SC like YBCO. A typical value of the 
London length $\lambda$ in the conduction planes $ab$ is $\lambda = 0.2 \ 
\mu m$. From the expression for the London length $\lambda = 
\sqrt{m^*/(\mu_0 n_s e^2)}$ one finds for the density $n_s$ of 
superconducting charge carriers $n_s \sim 10^{27} \ m^{-3}$. In a material 
with critical current density of the order of $10^8 \ A/m^2$, this 
corresponds to an intrinsic velocity of the carriers $v_j = j/\rho = 
j/(en_s) \sim 0.6 \ m/s$. For comparison, in a SC rotating at 100 $Hz$ 
(6000 $rpm$) the rotation velocity of the lattice 10 $cm$ apart from the 
axis is $v_{rot} \simeq 63 m/s$. Therefore, in this case the slide current 
can be up to 100 times larger than the critical intrinsic supercurrent.

\subsection*{3. London-Maxwell equations for accelerating SCs} 

In order to achieve a better understanding of the problem, it is useful to 
recall first in short the theoretical ingredients employed for the 
description of accelerated SCs at thermodinamic equilibrium. 

More generally, let us consider the interplay between a {\it moving} SC 
and the e.m.\ field. This includes the case of rotating SC samples, or 
samples which oscillate or accelerate along a line. Some of these cases 
were studied very early \cite{London,Becker}. A general formalism, 
suitable for the description of all these situations, has been given by 
Peng et al.\ \cite{Peng17,Peng}. They proposed a unified phenomenological 
approach to study the electrodynamics of both an arbitrary moving SC and a 
SC under the influence of non-electromagnetic external forces including 
the Newtonian gravitation and gravitational waves. This theoretical work 
has provided a basis for the analysis of several experiments which exploit 
the London field for precise determinations of the Cooper-pair mass and 
the ratio $\hbar/m$ \cite{Exps}; a similar analysis was applied to the 
readout systems of the Stanford gyroscope experiment \cite{Peng3}. 

Usually one starts from the covariant generalizations of the London or 
Ginzburg-Landau equations. The Ginzburg-Landau theory is needed if one 
wants to include non-linear effects and spatial variations in the order 
parameter $|\psi|^2$. If we focus on situations in which the perturbating 
fields and currents are so weak that $|\psi|^2 = n_s = const.$, then the 
Ginzburg-Landau equation reduces to the London equations which describe 
the motion of superelectrons. (In dealing with Type II SCs, we must 
further assume that the external magnetic field is zero, otherwise partial 
flux penetration will occur.) 

In order to account for the motion of the SC, London introduced the 
concept that the net current should be the sum of the supercurrent and the 
current due to the motion of the ions. He then combined the London 
equations, the equations of motion of ions, and Maxwell equations to study 
the electrodynamics of a rotating SC in the presence of e.m.\ fields. 
More generally, denoting by $u$ and $U$ the 4-velocities of superelectrons 
and ions, respectively, the net electric 4-current will be 
\begin{equation} 
J^\mu = 2n_s e(U^\mu-u^\mu) 
\label{e3} 
\end{equation} 

This equation must be added to the Maxwell equations and the covariant 
equations of motion for the superelectrons. In the non-relativistic limit 
of low velocities (${\bf v}$ for the electrons, ${\bf V}$ for the ions) 
these take the usual form 
\begin{eqnarray} 
& & \frac{d{\bf v}}{dt} \simeq -\frac{e}{m^*} {\bf E} + \frac{1}{m^*} 
{\bf f} \label{e4} \\ 
& & {\bf \partial} \times {\bf v} \simeq \frac{e}{m^*} {\bf B} + 
\frac{1}{m^*} \int dt \, {\bf \partial} \times {\bf f} \label{e5} 
\end{eqnarray} 
implying 
\begin{equation} 
{\bf v} = \frac{e}{m^*} {\bf A} + \frac{1}{m^*} \int dt \, {\bf f} 
\label{e6} 
\end{equation} 
where ${\bf f}$ is the external force acting on superelectrons. In 
conclusion, after taking the curl and time derivative, we find 
\begin{eqnarray} 
& & \partial^2 {\bf E} - \frac{d^2{\bf E}}{dt^2} = \frac{2\mu_0 n_s 
e^2}{m^*} \left( {\bf E} - \frac{{\bf f}}{e} + \frac{m^*}{e} \frac{d{\bf 
V}}{dt} \right) \label{e7} \\ 
& & \partial^2 {\bf B} - \frac{d^2{\bf B}}{dt^2} = \frac{2\mu_0 n_s 
e^2}{m^*} \left( {\bf B} + \frac{1}{e} \int dt \, {\bf \partial} 
\times {\bf f} - \frac{m^*}{e} {\bf \partial} \times {\bf V} \right) 
\label{e8} 
\end{eqnarray} 

By solving these equations with suitable boundary conditions and with the 
equations for the motion of ions and for the external forces acting on the 
superelectrons, one can describe the electrodynamics of an arbitrary 
moving SC in the presence of e.m.\ fields to lowest order. The forces 
acting on ions, which do not appear explicitly in these equations, are 
involved in the different expressions of ${\bf V}$ for various cases. We 
note, as stressed by Peng et al.\ \cite{Peng}, that: 

\medskip 

- an external force ${\bf f}$, the electric field and the acceleration of 
ions are coupled; 

\medskip 

- the curl of an external force, the magnetic field and the curl of ${\bf 
V}$ are coupled, too; 

\medskip 

- the motion of ions, ${\bf V}$, and the vector magnetic potential, play 
the same role. 

\medskip 

The effective electron mass $m^*$ coincides with the bare mass if the 
electrons are in relative equilibrium with the lattice (${\bf v}={\bf 
V}$), but in general it is different (the cyclotron frequency of charge 
carriers in a crystal depends on $m^*$---see for instance \cite{Kittel}). 

Note that the London equations (\ref{e4}), (\ref{e5})---a crucial 
component of the final eq.s (\ref{e7}), (\ref{e8})---are equivalent to the 
minimization of the free energy $F$ of the SC; thus they hold at 
thermodynamical equilibrium. One might wonder if transient phases can be 
important in the presence of sudden accelerations. For the cases to which 
these equations have been previously applied \cite{Peng}, they seem to 
work well, and this means that the system is always close to equilibrium. 
Layered SCs in fast non-uniform rotation may represent a notable 
exception. In the next section we describe a first heuristic approach to 
this problem. 

\subsection*{4. Heuristic description of non-equilibrium states} 

We have seen that a SC accelerating under the action of an external force 
can be described in a first approximation by the London-Maxwell equations 
(\ref{e7}), (\ref{e8}). All the quantities involved in these expressions 
are mutually coupled in a complex way, so the problem is hard to solve in 
general form. Moreover, in certain situations the system could be far from 
thermodynamic equilibrium. Let us then give here an heuristic description 
of the case of a rotating layered SC, taking advantage of the causal 
connections which are already known, namely: 

\medskip 

(i) a bulk ``slide current" due to a sudden acceleration generates, by 
virtue of the London equations (i.e., by minimization of the free energy 
of the SC), an increase in the surface supercurrent; 

\medskip 

(ii) by virtue of the Maxwell equations, the surface supercorrent produces 
a London field inside the SC; 

\medskip 

(iii) this field brings the Cooper pairs in relative equilibrium with the 
rotating lattice. 

\medskip 

Let us proceed by steps and illustrate three different possible 
situations: the case of an homogeneous rotating bar; the case of a 
rotating bar with two parts of comparable thickness, made of different 
materials; the case of a rotating bar with a thin layer of different 
material. 

\bigskip 

\noindent 
{\bf Case of an homogeneous rotating bar} 

\medskip 

Suppose an homogeneous cylindrical SC bar is rotating at angular velocity 
$\omega$ and equilibrium has been reached, with a London field $B_L = 
e\omega/m$ inside the bar. Then the bar accelerates, reaching an angular 
velocity $\omega'$. At first the superelectrons are left behind; therefore 
they recover at once their effective mass and a large bulk slide current 
arises. After that, however, the supercurrent at the surface of the bar 
grows, in such a way to produce a London field $B_L^* = 2\omega'm^*/e$. 
This sets the superelectrons again in equilibrium with the lattice, and 
finally (possibly after a few further ``oscillations" with respect to 
relative equilibrium) the field attains a new equilibrium value $B_L' = 
2\omega'm/e$. The whole process is fast and usually not observed in the 
classical experiments involving slow and steady rotors (compare our 
conclusions in Section 5). 

More precisely, note that in this situation the steps (i) and (ii) above 
can be best visualized through an analogy between the SC and an ideal 
solenoid, as follows: 

\medskip 

(i') A driving electromotive force (EMF) is applied to the solenoid. Like 
many EMFs, it is of ``thermodynamic" origin (minimization of the free 
energy of the SC) and takes a characteristic time $\Delta t_{EMF}$ to 
reach its maximum. 

\medskip 

(ii') The current in the solenoid grows in response to the external EMF, 
but due to the self-inductance of the system, this growth takes a 
characteristic time $\Delta t_{induction}$. 

\medskip 

The total time required to reach the new equilibrium configuration is thus 
of the order of $\Delta t_{equilibrium} = \Delta t_{EMF} + \Delta 
t_{induction}$. If this is much smaller than the characteristic 
acceleration time, then the system will just pass through a sequence of 
equilibrium states. 

In the following, however, we shall consider ``sudden accelerations", with 
characteristic times smaller than $\Delta t_{equilibrium}$.

\bigskip 

\noindent 
{\bf Case of a rotating bar with two parts of comparable thickness} 

\medskip 

Let us next consider a cylindrical bar made of two parts, 1 and 2, of 
comparable thickness (see figure, A). The material is different in the two 
parts, and so is the effective mass of the electrons. Suppose the system 
is initially in equilibrium: the bar rotates with constant angular 
velocity around its vertical axis and the London field is the same in both 
parts, corresponding to the bare electron mass 
($B_{L1}=B_{L2}=e\omega/m$). 

If the angular velocity is suddenly increased, a transient phase will 
follow. The skin supercurrents $j_1$ and $j_2$ must increase, too, in 
order to produce a stronger London field. In the meanwhile, the 
superelectrons are unable to follow the rotation frequency of the lattice, 
and are in relative motion with respect to it. 

The magnetic field needed to bring the electrons again to rest with 
respect to the lattice is different in the two parts, because the 
effective masses are different. The same is true for the skin 
supercurrents $j_1$ and $j_2$. Like in the previous case, there will be 
some oscillations around the relative equilibrium positions, but soon a 
new state is reached (if there are no further accelerations), with 
$B_{L1}'=B_{L2}'=e\omega'/m$. This is clearly the state with minimum total 
free energy $F=F_1+F_2$. 

\bigskip 

\noindent 
{\bf Case of a rotating bar with a thin layer of different material} 

\medskip 

Finally, we consider the previous case in the limit when the Part 2 of the 
cylindrical bar is much thinner than Part 1 (see figure). Let us also 
suppose that $m^*_2 > m^*_1$, i.e., the effective electron mass is larger 
in 1 than in 2. It is easy to see that after a sudden acceleration, the 
Part 2 cannot reach a new equilibrium situation, but remains in a sort of 
metastable state. 

In fact, following the acceleration the skin supercurrent $j_1$ increases 
until the London field $B_{L1}$ brings again the superelectrons in 1 in 
relative equilibrium with the lattice (possibly after some oscillations); 
this field, however, is not strong enough to establish relative 
equilibrium in 2, where the effective electron mass is larger. 

Note that being the Part 2 very thin (and much thinner than the radius of 
the bar), the magnetic field in it cannot be substantially different from 
$B_{L1}$. For the same reason, the free energy $F_2$ gives a negligible 
contribution to the total free energy, so while $F_1$ must be at a minimum 
and no bulk slide current can exist in 1, such a current can indeed be 
present in 2 in the circumstances we are considering. 

Also note that the ``feedback" magnetic field generated by this 
superficial current lies in a plane orthogonal to the bar axis; therefore 
it does not tend to compensate for the insufficient London field and does 
not oppose to the surface current. 

In practice, a thin layer like the Part 2 considered above can be present 
in a SC not only because of intentional differences in the oxigen doping 
of the material, but also for other reasons. For instance, the bulk of the 
material might have been subjected to a melting treatment, while the base 
was less affected because in contact with a coolant. 

The essential point, for the anomalous behavior described above to occur, 
is that the crystal structure of Part 2 must be different from that of 
Part 1, and the effective electron mass larger. This behavior could also 
depend on the temperature, because the different crystal structures of 
parts 1 and 2 could imply different critical temperatures, and typically 
we expect $T_{c2} < T_{c1}$. In this case, the metastable states will be 
most relevant at temperatures $T$ such that $T_{c2} < T < T_{c1}$.

\subsection*{5. Conclusions} 

In this work, after recalling in Section 3 the Maxwell-London equations 
for the general case of a SC in accelerated motion, we have set out in 
Section 4 an heuristic approach to the case of a layered SC in fast 
non-uniform rotation. 

We have seen that in this situation the system can enter non-equilibrium 
states. In particular, if one of the layers is much thinner than the 
others, and the effective electron mass in it larger, ``slide" surface 
currents can arise, with density higher than the critical density $j_c$ of 
the material. (Compare the estimate given in the Introduction, which yelds 
$j \sim 100 j_c$ for a 100\% difference between the rotation velocity of 
the lattice and that of the superconducting carriers; for smaller 
differences, $j/j_c$ varies in direct ratio). 

This phenomenon is interesting in itself, but also because in some 
experiments involving rotating SCs \cite{Noever} the real operating 
conditions are not far from those considered here in principle: namely one 
has large ceramic disks (10-30 $cm$ in diameter), rotating at frequencies 
of thousands of $rpm$, with acceleration and braking phases during which 
the rotation frequency varies by some \% in a few seconds; moreover, these 
large superconducting samples are often made by several layers, having 
different crystal structure and oxigen doping. (For comparison, the 
classical experiments \cite{Exps} involve rotors with a maximum size of 
$\sim 5 \ cm$ and maximum rotation frequencies of $\sim 5 \ Hz$, 
driven by steady gas flows). 

The possible presence, in these systems, of large surface currents like 
those predicted by our analysis, could be checked directly through 
magneto-optical techniques \cite{Vlasko}. Alternatively, one could look 
for indirect evidence, for instance investigating the effect of these 
anomalous currents on the material that supports them. 

\bigskip 

This work has been partially supported by the A.S.P., Associazione per lo 
Sviluppo Scientifico e Tecnologico del Piemonte, Turin - Italy.

\newpage 

\noindent FIGURE CAPTION 

\bigskip

\noindent Fig.\ 1 - A: rotating bar with two parts of comparable 
thickness; B: rotating bar with a thin layer of different material. The 
rotation axis is vertical in the figures. 


\begin{thebibliography}{99} 


\bibitem{London} F.\ London, {\it Superfluids}, Wiley, New York, 1950. 

\bibitem{Waldram} J.\ Waldram, {\it Superconductivity of metals and 
cuprates}, IoP, London, 1996. 

\bibitem{Liu} M.\ Liu, Phys.\ Rev.\ Lett.\ {\bf 81} (1998) 3223. 

\bibitem{Anderson} P.W.\ Anderson, in {\it Progress in Low Temperature 
Physics}, ed.\ C.J.\ Gorter, North Holland, Amsterdam, 1967. 

\bibitem{Gutfreund} B.\ Cabrera, H.\ Gutfreund and W.A.\ Little, Phys.\ 
Rev.\ {\bf B 25} (1982) 6644. 

\bibitem{Peskin} B.\ Cabrera and M.E.\ Peskin, Phys.\ Rev.\ {\bf B 39} 
(1989) 6425. 

\bibitem{Exps} S.B.\ Felch, J.\ Tate, B.\ Cabrera and J.T.\ Anderson, 
Phys.\ Rev.\ {\bf B 31} (1985) 7006; S.B.\ Felch, J.\ Tate and B.\ 
Cabrera, Phys.\ Rev.\ {\bf B 42} (1990) 7885. 

\bibitem{Becker} R.\ Becker, F.\ Sauter and C.\ Haller, Z.\ Phys.\ {\bf 
85} (1933) 772. 

\bibitem{Peng17} H.\ Peng, Sup.\ Sci.\ Tech.\ {\bf 3} (1990) 429; H.\ 
Peng, D.G.\ Torr, E.\ Hu and B.\ Peng, Phys.\ Rev.\ {\bf B 43} (1991) 
2700. 

\bibitem{Peng} H.\ Peng, G.\ Lind and Y.S.\ Chin, Gen.\ Rel.\ Grav.\ {\bf 
23} (1991) 1231. 

\bibitem{Peng3} J.D.\ Fairbank, B.S.\ Deaver, Jr., C.W.F.\ Everitt and 
P.F.\ Michelson, ed.s, {\it Near Zero: New Frontiers of Physics}, W.H.\ 
Freeman, New York, 1988; M.\ Cerdonio, G.\ Prodi and S.\ Vitale, Gen.\ 
Rel.\ Grav.\ {\bf 20} (1988) 83. 

\bibitem{Kittel} C.\ Kittel, {\it Introduction to solid state physics}, 
J.\ Wiley, New York, 1986, p.\ 196. 

\bibitem{Noever} N.\ Li, D.A.\ Noever, R.\ Koczor, T.\ Robertson, W.\ 
Brantley, Physica {\bf C 281} (1997) 260, and references; 
E.\ Podkletnov and R.\ Nieminen, Physica {\bf C 203} (1992) 441;
R.\ Koczor, D.\ 
Noever, {\it Fabrication of large YBCO superconducting disks}, report AIAA 
99-2147, June 1999. 

\bibitem{Vlasko} 
V.K.\ Vlasko-Vlasov, U.\ Welp, G.W.\ Crabtree, D.\ Gunter, V.V.\ 
Kabanov, V.I.\ Nikitenko, L.M.\ Paulius, Phys.\ Rev.\ {\bf B 58} 
(1998) 3446. 

\end{thebibliography}
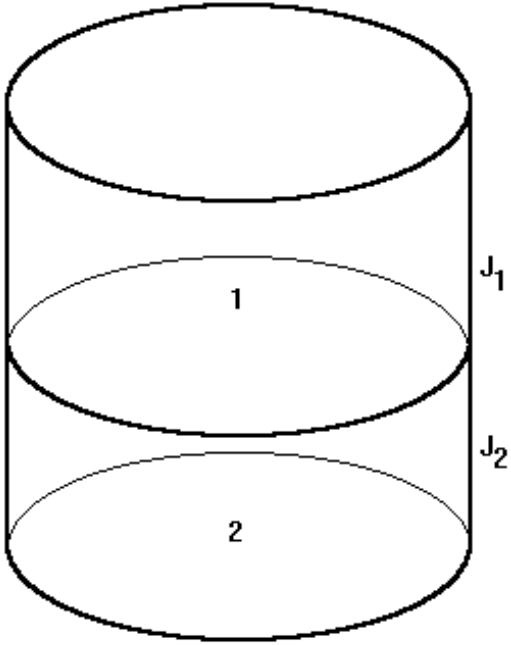
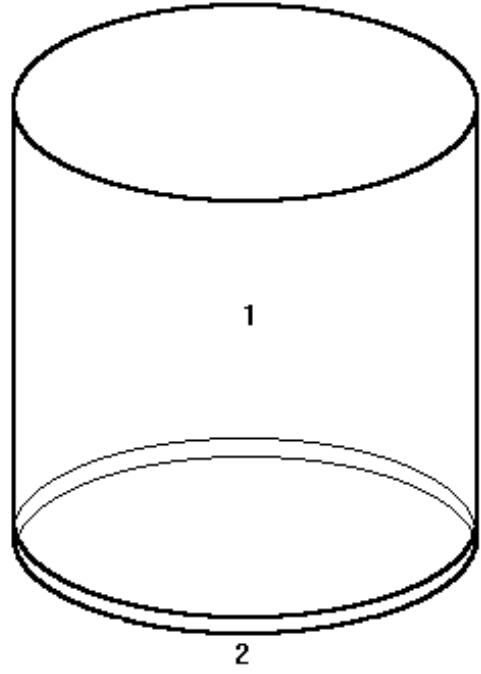
\end{document}